\documentclass[12pt]{article}
\usepackage{amsfonts}
\usepackage{bm}
\usepackage{graphicx}
\begin{document}

\title{
Quantifying geometric measure of entanglement by mean value of spin and spin correlations for pure and mixed states
}

\author{{\bf \sf A. M. Frydryszak$^1$, M. I. Samar$^2$, V. M. Tkachuk$^2$}\\[3mm]
 $^{1)}$ Institute of Theoretical Physics,\\
University of Wroclaw, pl. M. Borna 9,\\
50 - 204 Wroclaw, Poland\\
{\small \em e-mail: amfry@ift.uni.wroc.pl}
\\[2mm]
$^{2)}$ Department for Theoretical Physics,\\ Ivan Franko National
University of Lviv,\\ 12 Drahomanov St., Lviv, UA-79005, Ukraine\\
{\em \small  e-mail: voltkachuk@gmail.com}}

\maketitle

\begin{abstract}
We quantify the geometric measure of entanglement in terms of mean
values of observables of entangled system. For pure states we find
the relation of geometric measure of entanglement  with the mean
value of spin one-half for the system composed of spin  and
arbitrary quantum system. The geometric measure of entanglement
for mixed  states of rank-2 is
studied as well. We find the explicit expression for geometric
entanglement and the relation of entanglement in this case with
the values of spin correlations. These results allow to find
experimentally the value of entanglement by measuring a value of
the mean spin and the spin correlations for pure and mixed states,
respectively.
The obtained results are applied for calculation of entanglement during the evolution
in spin chain with Ising interaction , two-spin Ising model in transverse fluctuating magnetic
field, Schr\"odinger cat in fluctuating magnetic field.

\end{abstract}

\section{Introduction}

Quantification of entanglement is one of the principal challenges in quantum information theory
 \cite{Hor09,Ple07}.
Among the natural entanglement measures there is the geometric measure of entanglement proposed by Shimony
\cite{Shi95}. Its properties for multiqubit systems were studied by Brody and Hughston \cite{BroHu01} and also  Wei and Goldbart \cite{Wei03}.
A comparison of different definitions of the
geometric measure of entanglement can be found in \cite{Chen14}.

The geometric measure of entanglement is defined as a minimal squared distance between an entangled state
$| \psi\rangle$ and a set of
separable states $|\psi_s\rangle$
\begin{eqnarray} \label{DefE}
E(|\psi\rangle)=\min_{|\psi_s\rangle}(1-|\langle\psi|\psi_s\rangle|^2)=1-\max_{|\psi_s\rangle}|\langle\psi|\psi_s\rangle|^2,
\end{eqnarray}
where $1-|\langle\psi|\psi_s\rangle|^2$ is the squared distance of
Fubini-Study.
Note that
despite its simple definition it involves a nontrivial minimization
procedure over separable states.

In the case of mixed states the entanglement can be defined in terms of the convex roof construction
\begin{eqnarray}
E(\rho)={\rm min} \sum_i p_i E(|\psi_i\rangle),
\end{eqnarray}
where minimization is done over all possible decompositions of density matrix with respect to pure quantum states
\begin{eqnarray}\label{DM}
\rho=\sum_i p_i |\psi_i\rangle \langle\psi_i|, \quad\quad  \sum_i p_i=1.
\end{eqnarray}

 The essential question is what is the way to measure the entanglement directly.
 Many methods and schemes  were proposed for
 this purpose \cite{Guh02,Alt05,Kot07,Wal07,Enk07,Fei09,Bri10,Law14,Dai14,Bar15}.
In present paper we study geometric measure of entanglement for
pure and mixed states and find the relation of entanglement with
mean value observables. Namely, for pure states of a spin with
arbitrary quantum system we obtain exactly the relation of
the entanglement  with mean value of spin. In the case of
mixed states of rank-2 we find the explicit expression for
geometric entanglement  and the relation of the entanglement with
spin correlations for special cases of rank-2 mixed states. These
mean values  are experimentally measurable. Therefore, our results
give an additional possibility for  direct experimental
measurement of degree of entanglement.
From the other hand the obtained results give also
possibility to find in explicit form geometric entanglement for different quantum systems.

Note also that for mixed quantum states, even bipartite mixed
states of rank-2, many questions remain opened (see for instance
\cite{Boy16} and references therein). Therefore study of entanglement in
mixed states of rank-2 remains interesting and actual.

This paper is organized as follows: In Section 2 we find the relation of geometric entanglement of spin with arbitrary quantum system in pure state with mean value of spin. In Section 3 we study
the geometric measure of entanglement for rank-2 mixed states and find relation of entanglement
with spin correlations. In section 4 we apply the results obtained in section 2 and 3 for calculation  of geometric entanglement during the pure evolution in spin chin and for calculation  of geometric entanglement during
the evolution of ensemble of two-spin systems in fluctuating magnetic field and for calculation of entanglement
during the evolution and the decoherence of Schr\"odinger cat.
And finally, the conclusions are presented in Section 5. The minimization
procedure over separable states for geometric entanglement in mixed state is presented in Appendix.

\section{Characterizing entanglement of spin with arbitrary quantum system by mean value of spin}

In general, the pure quantum state of spin one-half (or qubit) which can be entangled with other arbitrary  quantum system in pure state reads

\begin{eqnarray}\label{PsiG}
|\psi\rangle=a|\uparrow\rangle|\phi_1\rangle+b|\downarrow\rangle|\phi_2\rangle,
\end{eqnarray}
here $|\phi_1\rangle$ and $|\phi_2\rangle$ are arbitrary
state vectors of quantum system entangled with a spin, constants $a,b$ are
real and positive, phase multipliers can be included into
$|\phi_1\rangle$ and $|\phi_2\rangle$, which satisfy
normalization conditions
$\langle\phi_1|\phi_1\rangle=\langle\phi_2|\phi_2\rangle=1$. Note
that in general this functions are not orthogonal
$\langle\phi_1|\phi_2\rangle\ne 0$.

Arbitrary
state vector of spin, interacting with some quantum system, can be represented by the Schmidt decomposition
\begin{eqnarray}\label{PsiG}
|\psi\rangle=\lambda_1|\alpha_1\rangle|\tilde\phi_1\rangle+\lambda_2|\alpha_2\rangle|\tilde\phi_2\rangle,
\end{eqnarray}
where $|\alpha_1\rangle$, $|\alpha_2\rangle$ are two orthogonal states of spin
\begin{eqnarray}
|\alpha_1\rangle={|\uparrow\rangle +\alpha|\downarrow\rangle\over \sqrt{1+|\alpha|^2}}, \ \
|\alpha_2\rangle={{\alpha}^{*}|\uparrow\rangle -|\downarrow\rangle\over \sqrt{1+|\alpha|^2}},
\end{eqnarray}
and $|\tilde\phi_1\rangle$, $|\tilde\phi_2\rangle$ are two orthogonal states of arbitrary quantum system interacting with the spin, $\langle\tilde\phi_1|\tilde\phi_2\rangle=0$. Constants $\lambda_1,\lambda_2$ are real and positive satisfying normalization condition $\lambda_1^2+\lambda_2^2=1$.

The geometric measure of entanglement is related with maximum value of squared Schmidt coefficients ($\lambda_1,\lambda_2$) \cite{Sen10} namely,
\begin{eqnarray}\label{EntSchm}
E(|\psi\rangle)=1-{\rm max}(\lambda_1^2,\lambda_2^2).
\end{eqnarray}

It turns out  that Schmidt coefficients $\lambda_1, \lambda_2$  are related with the mean value of spin. To show this, let us calculate squared mean value of spin
\begin{eqnarray} \label{sigmaSm}
\left<\bm{\sigma}\right>^2=
(\lambda_1^2-\lambda_2^2)^2=(1-2\lambda_1^2)^2=(1-2\lambda_2^2)^2.
\end{eqnarray}
From this relation we have
\begin{eqnarray}
\lambda^2_{1,2}={1\over2}(1\pm|\left<\bm{\sigma}\right>|),
\end{eqnarray}
where $|\left<\bm{\sigma}\right>|=\sqrt{\left<\bm{\sigma}\right>^2}$.
Hence,  the geometric measure of entanglement given by (\ref{EntSchm}) reads
\begin{eqnarray}\label{EntSpin}
E(|\psi\rangle)={1\over2}\left(1-|\left<\bm{\sigma}\right>|\right).
\end{eqnarray}
The entanglement of spin with other quantum system is entirely determined by the mean value of spin.
Note that we do not need an explicit expression for the Schmidt decomposition in order to calculate the entanglement
using (\ref{EntSpin}). It is only  important that this decomposition exists.

When a spin state
$|\chi\rangle$ is separable from a state of other system $ |\phi\rangle$
\begin{eqnarray}
|\psi\rangle= |\chi\rangle|\phi\rangle, \ \ |\chi\rangle=a|\uparrow\rangle+b|\downarrow\rangle
\end{eqnarray}
then
\begin{eqnarray}
\left<\bm{\sigma}\right>^2=\left<\chi|\bm{\sigma}|\chi\right>^2=1.
\end{eqnarray}
Thus, in this case $E=0$, as expected.
Note also, that maximal entanglement of spin system with other quantum system is achieved for configuration with vanishing mean value of spin,
$\left<\bm{\sigma}\right>^2=0$. As follows from (\ref{EntSpin}) spin and quantum system are separable when $|\left<\bm{\sigma}\right>|=1$.

So, we can
establish the value of entanglement of spin with other quantum system by measuring local properties of quantum system in pure state, namely mean value of spin.

\section{Entanglement of rank-2 mixed states and its relation to mean value of spin correlations}

We consider special cases of mixed states of two spins. The first one is the case of mixed states with density matrix (\ref{DM})
where $|\psi_i\rangle$ are given on subspace spanned
by vectors $|\uparrow\downarrow\rangle$, $|\uparrow\downarrow\rangle$.
The second one is the case of mixed states with density matrix (\ref{DM})
where $|\psi_i\rangle$ are given on subspace spanned
by vectors $|\uparrow\uparrow\rangle$, $|\downarrow\downarrow\rangle$.
For these two cases of rank-2 mixed states we are able to express entanglement over the mean value of spin correlations.

Let us consider in details the first case for which
arbitrary vector of pure state
can be written in the form similar to spin-$1/2$ state vector
\begin{eqnarray} \label{psiFirst}
|\psi\rangle=\cos {\theta\over2} |\Uparrow\rangle +\sin {\theta\over2}e^{i\phi}|\Downarrow\rangle,
\end{eqnarray}
where we introduce the notation
\begin{eqnarray}
|\Uparrow\rangle=|\uparrow\downarrow\rangle, \ \ |\Downarrow\rangle=|\downarrow\uparrow\rangle.
\end{eqnarray}
Moreover we can introduce the analog of Pauli operators
acting on this subspace
\begin{eqnarray}\label{SIGMA}
\Sigma^x=\sigma_1^x\sigma_2^x, \ \ \Sigma^y=\sigma_1^y\sigma_2^x, \ \ \Sigma^z=\sigma_1^z\sigma_2^0,
\end{eqnarray}
where $\sigma_i^{\alpha}$ Pauli operators for spin $i$.
Note, we use traditional notation i.e.
when dedicated index enumerates different systems we assume tensor product
e.g. $\sigma_1^x\sigma_2^x=\sigma_1^x\otimes\sigma_2^x$.
One can verify that $\Sigma^x$, $\Sigma^y$, $\Sigma^z$ satisfy all properties of Pauli matrices and act on
$|\Uparrow\rangle$, $|\Downarrow\rangle$ in the same way as Pauli matrices act on $|\uparrow\rangle$, $|\downarrow\rangle$. Matrix representation of introduced operators (\ref{SIGMA}) on subspace $|\Uparrow\rangle$, $|\Downarrow\rangle$ reads
\begin{eqnarray}
\Sigma^x=\left(
  \begin{array}{cc}
    0 & 1 \\
    1 & 0 \\
  \end{array}
\right),
\Sigma^y=\left(
  \begin{array}{cc}
    0 & -i \\
    i & 0 \\
  \end{array}
\right),
\Sigma^z=\left(
  \begin{array}{cc}
    1 & 0 \\
    0 & -1 \\
  \end{array}
\right).
\end{eqnarray}

Density matrix for a pure quantum state (\ref{psiFirst}) can be written in the form
\begin{eqnarray}\label{DMP}
\rho={1\over 2}(1+{\bf a}\cdot{\bf \Sigma}),
\end{eqnarray}
where ${\bf a}$ plays the role of Bloch vector with $|{\bf a}|=1$, direction of this vector is given by spherical angles $\theta$ and $\phi$.

Density matrix of mixed state for considered subspace reads
\begin{eqnarray}\label{DMM}
\rho=\sum_i p_i \rho_i={1\over 2}(1+\sum_ip_i{\bf a_i}\cdot{\bf \Sigma}),
\end{eqnarray}
where $\rho_i$ is density matrix corresponding to a pure state
\begin{eqnarray} \label{psiFirsti}
|\psi_i\rangle=\cos {\theta_i\over2} |\Uparrow\rangle +\sin {\theta_i\over2}e^{i\phi_i}|\Downarrow\rangle,
\end{eqnarray}
unit vector ${\bf a_i}$ is defined by angles $\theta_i$ and $\phi_i$.
Density matrix of mixed state (\ref{DMM})
has the same form as pure state density matrix (\ref{DMP}) with
\begin{eqnarray}\label{adecomp}
{\bf a}=\sum_ip_i{\bf a_i},
\end{eqnarray}
where $\bf a$ entirely determines density matrix, $ |{\bf a}|\le 1$ correspond to mixed states and $|{\bf a}|=1$
corresponds to pure ones.

According to (\ref{EntSpin}) the geometric entanglement of one spin with other in pure state $|\psi_i\rangle$ reads
\begin{eqnarray}\label{EntSpini}
E(|\psi_i\rangle)={1\over2}\left(1-|\left<\psi_i|\bm{\sigma_1}|\psi_i\right>|\right).
\end{eqnarray}
It can be seen that for an arbitrary two-spin quantum state  we have $\left<\bm{\sigma_1}\right>^2=\left<\bm{\sigma_2}\right>^2$. So, in (\ref{EntSpini}) we can use mean value of first spin or mean value of the second one. It means that the measure of entanglement of one spin with an other one is symmetric with respect to the spin subsystems.

One can find that for state (\ref{psiFirsti}) mean values of projections of the first spin are
\begin{eqnarray}
\left<\psi_i|{\sigma_1^x}|\psi_i\right>=0, \ \ \left<\psi_i|{\sigma_1^y}|\psi_i\right>=0, \ \
\left<\psi_i|{\sigma_1^z}|\psi_i\right>=\cos\theta_i=a_i^z.
 \end{eqnarray}
 Therefore, the geometric measure of entanglement of two spins in pure state (\ref{psiFirsti}) is
\begin{eqnarray}\label{EntSpiniZ}
E(|\psi_i\rangle)={1\over2}\left(1-|\left<\psi_i|{\sigma^z_1}|\psi_i\right>|\right)=
{1\over2}\left(1-|a_i^z|\right)
\end{eqnarray}
and geometric entanglement of two spins in mixed state has the following form
\begin{eqnarray}
E(\rho)={1\over2}\left(1-\max\sum p_i|a_i^z|\right).
\end{eqnarray}

Now we have the problem to find $\max p_i\sum|a_i^z|$ over all decomposition of density matrix, namely, decomposition of fixed ${\bf a}$ over ${\bf a_i}$ and $p_i$ according to (\ref{adecomp}).
Note that $|{\bf a_i}|=1$ and therefore
\begin{eqnarray}
\sum_i p_i |{\bf a_i}|=1.
\end{eqnarray}
It is convenient to introduce the vector ${\bf l_i}=p_i {\bf a_i}$ in terms of which we look for
\begin{eqnarray}\label{propl}
\max\sum_i |l_i^z|
\end{eqnarray}
under condition
\begin{eqnarray}\label{cond}
\sum_i {\bf l_i}= {\bf a},
\end{eqnarray}
with constraint
\begin{eqnarray}\label{const}
\sum_i |{\bf l_i}|=1.
\end{eqnarray}
Interestingly enough that this problem has geometric interpretation which is useful for its solving (for explicit construction cf. the Appendix). As  a result we obtain
\begin{eqnarray}
\max\sum_i |l_i^z|=\sqrt{1-a^2\sin^2\theta},
\end{eqnarray}
where
$a=|{\bf a}|$, $\theta$ is angle between $\bf a$ and $z$ axis. Because $\max p_i\sum|a_i^z|=\max\sum_i |l_i^z|$, finally
we get
\begin{eqnarray}\label{Erhoa}
E(\rho)={1\over 2}\left(1-\sqrt{1-a^2\sin^2\theta}\right)={1\over 2}\left(1-\sqrt{1-a^2_x-a^2_y}\right).
\end{eqnarray}

To express this result by the mean value of spin correlations,  note that the components of the Bloch vector
\begin{eqnarray}
a^x=<\Sigma^x>=<\sigma_1^x\sigma_2^x>, \\
a^y=<\Sigma^y>=<\sigma_1^y\sigma_2^x>, \\
a^z=<\Sigma^z>=<\sigma_1^z>,
\end{eqnarray}
where $<A>={\rm Sp} A\rho$ is mean value for mixed state.
Thus we have
\begin{eqnarray} \label{Ecorrel}
E(\rho)={1\over 2}\left(1-\sqrt{1-<\sigma_1^x\sigma_2^x>^2-<\sigma_1^y\sigma_2^x>^2}\right).
\end{eqnarray}
Thus, the geometric entanglement of two spins in mixed states
formed on subspace spanned by vectors $|\uparrow\downarrow\rangle$, $|\uparrow\downarrow\rangle$
can be written in terms of the mean values of spin correlations.

One can verify that result given by (\ref{Ecorrel}) is valid also in the case of
mixed states on subspace spanned by vectors
\begin{eqnarray}
|\Uparrow\rangle=|\uparrow\uparrow\rangle, \ \ |\Downarrow\rangle=|\downarrow\downarrow\rangle.
\end{eqnarray}

Analyzing (\ref{Erhoa}) we can conclude that the maximally entangled states with $E=1/2$ can be obtained only in the case of $|{\bf a}|=1$, which corresponds to pure states. An arbitrary mixed state has magnitude of entanglement less than the maximal value $1/2$.
From (\ref{Ecorrel}) follows that two-spin states in considered family of mixed states are non entangled when
$<\sigma_1^x\sigma_2^x>=0$ and $<\sigma_1^y\sigma_2^x>=0$.

The obtained result can be generalized for rank-2 nixed state of arbitrary number of spins $N$.
Let us consider the rank-2 nixed state of $N$ spins with density matrix (\ref{DM})
where $|\psi_i\rangle$ are given on subspace spanned
by vectors
\begin{eqnarray}
|\Uparrow\rangle=|\uparrow\uparrow...\uparrow\rangle, \ \ |\Downarrow\rangle=|\downarrow\downarrow...\downarrow\rangle.
\end{eqnarray}
Now similarly to (\ref{SIGMA}) we can introduce the following analog of Pauli operators
acting on this subspace
\begin{eqnarray}\label{SIGMAN}
\Sigma^x=\sigma_1^x\sigma_2^x...\sigma_N^x, \ \ \Sigma^y=\sigma_1^y\sigma_2^x... \sigma_N^x,\ \ \Sigma^z=\sigma_1^z\sigma_2^0...\sigma_N^0.
\end{eqnarray}
Therefore geometric entanglement (\ref{Erhoa}) derived for two spins in mixed state is suitable also for the case of $N$ spins.
Difference is only that $\Sigma$ operators now is given by (\ref{SIGMAN}) instead of (\ref{SIGMA}).
Note that in the case of $N$ spin (\ref{Erhoa}) describe the geometric entanglement of first spin with others.
When we wont to find geometric entanglement, for instance, second spin with others then operators (\ref{SIGMAN}) must be changed to
\begin{eqnarray} \label{SIGMAN2}
\Sigma^x=\sigma_1^x\sigma_2^x\sigma_3^x...\sigma_N^x, \ \ \Sigma^y=\sigma_1^x\sigma_2^y\sigma_3^x... \sigma_N^x,\ \ \Sigma^z=\sigma_1^0\sigma_2^z\sigma_3^0...\sigma_N^0.
\end{eqnarray}
In the same way we can find entanglement of arbitrary spin with others.

\section{Calculation of entanglement in particular cases of spin systems}
\subsection{Entanglement in spin chain. Pure states.}
In this section we study the entanglement of one spin with others in a spin chain. Relation (\ref{EntSpin}) between
entanglement and mean value of spin turns out to be useful for this purpose.
Let us consider the spin chain with Ising Hamiltonian
\begin{eqnarray}\label{H}
H=J\sum_{i=1}^{N-1}\sigma_i^x\sigma_{i+1}^x,
\end{eqnarray}
where $N$ is the number of spins in chain, $\sigma_i^x$ is the Pauli matrix of $i$-th spin.
We consider the evolution of spins starting at time $t=0$ from a factorized  state with zero entanglement
\begin{eqnarray}\label{psi0}
|\psi_{t=0}\rangle=|\psi_1\rangle|\psi_2\rangle\cdots |\psi_N\rangle,
\end{eqnarray}
where
\begin{eqnarray}
|\psi_i\rangle=a_i|\uparrow\rangle_i+b_i|\uparrow\rangle_i
\end{eqnarray}
is the state of $i$-th spin.

Interaction with Hamiltonian (\ref{H}) leads to the appearance of entanglement during
the evolution.
We study the entanglement of the first spin with other $N-1$ spins at time $t$.
In order to calculate the magnitude of the entanglement we use formula (\ref{EntSpin})
relating entanglement with mean value of spin.
In our case it is necessary to calculate the mean value of the first spin, namely
\begin{eqnarray} \label{ms}
\left<\bm{\sigma}_1\right>=\left<\psi(t)|\bm{\sigma}_1|\psi(t)\right>,
\end{eqnarray}
where vector of state at time $t$ is given by
\begin{eqnarray}
|\psi(t)\rangle=\exp(-i\omega t \sum_{i=1}^{N-1}\sigma_i^x\sigma_{i+1}^x)|\psi_{t=0}\rangle
=\prod_{i=1}^N\exp(-i\omega t \sigma_i^x\sigma_{i+1}^x)|\psi_{t=0}\rangle,
\end{eqnarray}
here $\omega=J/\hbar$. Substituting it into (\ref{ms}) we find that exponents in the operator of evolution which does not contain $\sigma_1^x$ is canceled.
As a result, for the mean value of the first spin we obtain
\begin{eqnarray}
\left<\bm{\sigma}_1\right>=\left<\psi_2|\left<\psi_1|e^{i\omega t\sigma_1^x\sigma_{2}^x}\bm{\sigma}_1e^{-i\omega t\sigma_1^x\sigma_{2}^x}|\psi_1\right>|\psi_2\right>
\end{eqnarray}
with components
\begin{eqnarray}
\langle\sigma_1^x\rangle=\langle\sigma_1^x\rangle_0,\\
\langle\sigma_1^y\rangle=\cos2\omega t \langle\sigma_1^y\rangle_0-\sin2\omega t \langle\sigma_1^z\rangle_0\langle\sigma_2^x\rangle_0,\\
\langle\sigma_1^z\rangle=\cos2\omega t \langle\sigma_1^z\rangle_0+\sin2\omega t \langle\sigma_1^y\rangle_0\langle\sigma_2^x\rangle_0,
\end{eqnarray}
where $\langle\sigma_i^{\alpha}\rangle_0=\langle\psi_i|\sigma_i^{\alpha}|\psi_i\rangle_0$ is the mean value of $i$-th spin ($i=1,2$, $\alpha=x,y,z$) in the initial state at $t=0$.
Then according to (\ref{EntSpin}) the geometric entanglement of the first spin with others in the spin chain reads
\begin{eqnarray}\label{Echain}
E={1\over2}\left(1-\sqrt{\langle\sigma_1^x\rangle_0^2+(\cos^22\omega t+\sin^22\omega t\langle\sigma_2^x\rangle_0^2)
(\langle\sigma_1^y\rangle_0^2+\langle\sigma_1^z\rangle_0^2)}\right).
\end{eqnarray}
It is interesting to note that the entanglement of the first spin with the rest
of the spin chain depends only on the mean value of the first and second
spins that is the result of nearest-neighbor interactions in Hamiltonian. One can verify that at $t=0$
the entanglement is zero as it must be for factorized state. Really, at $t=0$ under the square root we have $\left<\bm{\sigma}_1\right>_0^2$ that is equal to unity for an arbitrary state of the first spin and therefore $E=0$ for the initial state.

Now let us apply (\ref{Echain}) for some concrete initial states. Let the state
\begin{eqnarray} \label{psix}
|\psi_i\rangle={1\over\sqrt2}\left(|\uparrow\rangle_i\pm|\uparrow\rangle_i\right)
\end{eqnarray}
is the eigenstate of $\sigma_i^x$. In this case the initial state (\ref{psi0}) is the eigenstate of Hamiltonian (\ref{H}).
Therefore, the initial state does not change during the evolution and thus entanglement for all times is zero.
One can verify that the same result follows from (\ref{Echain}). For (\ref{psix}) the mean value of the components for the first spin are $\langle\sigma_1^x\rangle_0=\pm1$, $\langle\sigma_1^y\rangle_0=\langle\sigma_1^z\rangle_0=0$ and according to (\ref{Echain}) in this case $E=0$.

Notice that $E=0$ when only the second spin is in state (\ref{psix}). Then
$\langle\sigma_2^x\rangle_0^2=1$ and under the square root we have $\left<\bm{\sigma}_1\right>_0^2$ that is equal to unity for an arbitrary state of the first spin and therefore $E=0$.
Thus, in order to generate the entanglement between first
spin and others the mean value of $x$-component of the second spin in the initial state must satisfy condition $\left<\sigma_2^x\right>_0^2\ne 1$.

Now let us consider the initial state for $N$ spins
\begin{eqnarray}
|\psi_{t=0}\rangle=|\uparrow\rangle_1|\uparrow\rangle_2\cdots |\uparrow\rangle_N.
\end{eqnarray}
In this case
\begin{eqnarray}\label{EN}
E={1\over2}\left(1-|\cos 2\omega t|\right).
\end{eqnarray}

 Finally let us stress that the relation between entanglement and mean value of spin (\ref{EntSpin}) plays the crucial role in the calculation of the entanglement during the evolution of spins. As result it is not necessary to find state vector during the evolution explicitly. We can directly calculate the mean value of spin
and determine the entanglement.

\subsection{Entanglement of two spins in fluctuating magnetic field. Mixed states.}
In this Section we demonstrate the usefulness of formula (\ref{Erhoa}) for calculation entanglement of two spins in mixed state. For this purpose we consider ensemble of two-spin systems described by Ising Hamiltonian and placed in transverse fluctuating magnetic field
\begin{eqnarray} \label{TrIsing}
H=B(\sigma_1^z+\sigma_2^z)+J\sigma_1^x\sigma_2^x,
\end{eqnarray}
here $B$ is magnetic field. Note that magnetic fields of different magnitudes are applied
to different two-spin systems from this ensemble.

Hamiltonian (\ref{TrIsing}) has two invariant subspaces.
First one is spanned
by vectors $|\uparrow\uparrow\rangle$, $|\downarrow\downarrow\rangle$
and second one is
spanned by vectors $|\uparrow\downarrow\rangle$, $|\uparrow\downarrow\rangle$. Eigenvectors of
(\ref{TrIsing}) belong to these subspaces.

We consider the following problem. Let at the initial time $t=0$ all pairs of spins from the ensemble are in the same separated state
\begin{eqnarray}
|\psi_{t=0}\rangle=|\uparrow\rangle_1|\uparrow\rangle_2=|\uparrow\uparrow\rangle.
\end{eqnarray}
Vector of evolution in this case belongs to the subspace spanned
by vectors $|\uparrow\uparrow\rangle$, $|\downarrow\downarrow\rangle$.
As result of evolution of different two-spin systems in different magnetic fields we obtain mixed state.
Our goal is to find entanglement of two spins in this mixed state.

One can also easy verify that for this subspace we have
\begin{eqnarray}\label{H2}
H^2=J^2+4B^2=\hbar^2(\omega^2+\Omega^2),
\end{eqnarray}
here for the convenience we introduce the notations
\begin{eqnarray} \label{BJ}
B={\hbar\Omega\over 2}, J=\hbar\omega.
\end{eqnarray}

As result of (\ref{H2}) evolution operator can be written in the following form
\begin{eqnarray}\label{Ev}
e^{-iHt/\hbar}=\cos{\Omega_0 t}-i{H\over\hbar\Omega_0}\sin{\Omega_0 t},
\end{eqnarray}
where
$\Omega_0=\sqrt{\omega^2+\Omega^2}$.
Then, the evolution of two-spin system from the ensemble reads
\begin{eqnarray}\label{psitwo}
|\psi(t)\rangle=e^{-iHt/\hbar}|\uparrow\uparrow\rangle=(\cos\Omega_0t-i{\Omega\over\Omega_0}\sin\Omega_0 t)|\uparrow\uparrow\rangle
-i{\omega\over\Omega_0}\sin\Omega_0 t|\downarrow\downarrow\rangle.
\end{eqnarray}
The evolution of ensemble of two-spin systems is described by density matrix
\begin{eqnarray}\label{rhotwo}
\rho=\int d\Omega P(\Omega)|\psi(t)\rangle\langle\psi(t)|,
 \end{eqnarray}
where $P(\Omega)$ is distribution function of magnitude of magnetic field, $\Omega$ is related with magnetic field by (\ref{BJ}).
Substituting (\ref{psitwo}) into (\ref{rhotwo}) and using $|\uparrow\uparrow\rangle$ and
$|\downarrow\downarrow\rangle$ as basis vectors,  we find density matrix describing evolution of ensemble of two-spin systems in fluctuating magnetic field
\begin{eqnarray}
&&\hspace*{-0.6cm}\rho=\\
&&\hspace*{-0.6cm}=\left(
  \begin{array}{cc}
    \langle\cos^2\Omega_0t\rangle_{\Omega}+\langle{\Omega^2\over\Omega^2_0}\sin^2\Omega_0t\rangle_{\Omega} & {i\over2}\langle{\omega\over\Omega_0}\sin2\Omega_0t\rangle_{\Omega}+ \langle{\Omega\omega\over\Omega^2_0}\sin^2\Omega_0t\rangle_{\Omega}\\
    -{i\over2}\langle{\omega\over\Omega_0}\sin2\Omega_0t\rangle_{\Omega}+ \langle{\Omega\omega\over\Omega^2_0}\sin^2\Omega_0t\rangle_{\Omega} & \langle{\omega^2\over\Omega^2_0}\sin^2\Omega_0t\rangle_{\Omega} \\
  \end{array}
\right),\nonumber
\end{eqnarray}
where
$<f(\Omega)>_{\Omega}=\int d\Omega P(\Omega)f(\Omega)$.
This density matrix can be written in form (\ref{DMP}) where
\begin{eqnarray}
a_x=2\langle{\Omega\omega\over\Omega^2_0}\sin^2\Omega_0t\rangle_{\Omega},\\
a_y=-\langle{\omega\over\Omega_0}\sin2\Omega_0t\rangle_{\Omega},\\
a_z=1-2\langle{\omega^2\over\Omega^2_0}\sin^2\Omega_0t\rangle_{\Omega}.
\end{eqnarray}
Substituting this result into (\ref{Erhoa}) we find explicit expression for geometric entanglement
\begin{eqnarray}\label{Ent1}
E={1\over 2}\left(1-\sqrt{1-4\langle{\Omega\omega\over\Omega^2_0}\sin^2\Omega_0t\rangle^2_{\Omega}
-\langle{\omega\over\Omega_0}\sin2\Omega_0t\rangle^2_{\Omega}}\right).
\end{eqnarray}

Let us consider the following distribution function
\begin{eqnarray}\label{D1}
P(\Omega)={1\over2}\left(\delta(\Omega-\chi)+\delta(\Omega+\chi)\right).
\end{eqnarray}
It means that magnetic field has the same magnitude $|B|={\hbar\chi/2}$, the direction of magnetic field is change only. Namely, with probability $1/2$
magnetic field has positive direction along $z$-axis and with the same probability negative direction along $z$-axis.
As a result, the mean value of magnetic field is zero, $\chi$ characterize the value of fluctuation of magnetic field.
In this case geometric entanglement reads
\begin{eqnarray}\label{Emix+-}
E={1\over 2}\left(1-\sqrt{1
-{\omega^2\sin^22\sqrt{\chi^2+\omega^2}t\over \chi^2+\omega^2}}\right).
\end{eqnarray}

In the case when fluctuation of magnetic field is zero, $\chi=0$, the model considered in this Section corresponds to the model considered in the previous Section 4.1
for number of spins $N=2$.
One can verify that for $\chi=0$ equation (\ref{Emix+-}) reproduce (\ref{EN}) as it must be.
Note that increasing of fluctuation of magnetic field leads to the decreasing of entanglement. In the limit
$\chi\to \infty$ when fluctuations of magnetic field tend to infinity the geometric entanglement tends to zero.

Now let us consider Gausian distribution function
\begin{eqnarray}\label{D2}
P(\Omega)={\tau\over\sqrt{\pi}}e^{-\tau^2\Omega^2}.
\end{eqnarray}
In this case for large time we find the following asymptotic for geometric entanglement
\begin{eqnarray}
E={\omega\tau^2\over 4t}\sin^2(2\omega t+{\pi\over4}), \ \ t\to \infty,
\end{eqnarray}
that tends to $0$ when time $t\to \infty$.
Note that for distribution function (\ref{D1}) the entanglement is periodic function in time but for
Gausian distribution function (\ref{D2}) the entanglement tends to zero when time go to infinity.
Thus, the behavior of entanglement in time for mixed states essentially depends on
distribution function of fluctuating magnetic fields.

Finally
let us analyze evolution of system under consideration, starting from the initial state
\begin{eqnarray}
|\psi_{t=0}\rangle=|\uparrow\rangle_1|\downarrow\rangle_2=|\downarrow\uparrow\rangle.
\end{eqnarray}
In this case the vector of evolution belongs to the second subspace spanned by vectors $|\uparrow\downarrow\rangle$, $|\uparrow\downarrow\rangle$.
Hamiltonian in this subspace satisfies the following relation
\begin{eqnarray}
H^2=\omega^2.
\end{eqnarray}
Operator of evolution in this case has the form (\ref{Ev}) where instead of $\Omega_0$ we have $\omega$.
Note also that action of operator $B(\sigma_1^z+\sigma_2^z)$ on $|\uparrow\downarrow\rangle$ or $|\uparrow\downarrow\rangle$
is zero. Therefore fluctuating magnetic field has not influence on the evolution which is now described by pure state.
Making similar calculation as in the first case for entanglement we obtain the result  (\ref{Ent1}) where
$\Omega=0$ and $\Omega_0$ is changed to $\omega$. As a result for geometric entanglement we obtain the same formula
as in  (\ref{EN}).

\subsection{Decoherence of Schr\"odinger cat and geometric entanglement}
Let us consider N-spin systems placed in fluctuating magnetic field  with hamiltonian
\begin{eqnarray}
H= \sum_i^N {\hbar\Omega_i\over2}\sigma_i^z.
\end{eqnarray}
We suppose that distribution function for magnetic fields acting on different spins is independent
\begin{eqnarray}
P(\Omega_1,\Omega_2,...\Omega_N)=P(\Omega_1)P(\Omega_2)...P(\Omega_N)
\end{eqnarray}

In the initial time $t=0$ the system is in pure Schr\"odinger cat quantum state
\begin{eqnarray}
|\psi_{t=0}\rangle={1\over\sqrt2}(|\Uparrow\rangle+|\Downarrow\rangle).
\end{eqnarray}
One can easily find the density matrix describing the evolution of Schr\"odinger cat in fluctuating magnetic field
\begin{eqnarray}
\rho={1\over2}\left(
       \begin{array}{cc}
         1 & \langle e^{-i\Omega t}\rangle_{\Omega}^N \\
         \langle e^{i\Omega t}\rangle_{\Omega}^N & 1 \\
       \end{array}
     \right)=
     {1\over2}\left(
       \begin{array}{cc}
         1 &  e^{-Nt^2/4\tau^2} \\
         e^{-Nt^2/4\tau^2} & 1 \\
       \end{array}
     \right).
\end{eqnarray}
where we write the explicit expression for density matrix in the case
of Gausian distribution function (\ref{D2}). The decoherence in time is more quicker for larger $N$.
From (\ref{Erhoa}) we find explicit expression
for geometric entanglement of one spin with others in this case
\begin{eqnarray}
E(\rho)={1\over 2}\left(1-\sqrt{1-e^{-Nt^2/2\tau^2}}\right).
\end{eqnarray}
Thus decoherence leads to decreasing of the entanglement to zero.
\section{Conclusions}

In this paper we have studied the geometric measure of entanglement of spin-$\frac{1}{2}$ with other quantum system for pure and mixed states.
The main result of the present paper is given by (\ref{EntSpin}) for pure states and (\ref{Erhoa}) or(\ref{Ecorrel}) for mixed states.

In the case of pure quantum states we have shown that entanglement is entirely determined by the mean value of spin (\ref{EntSpin}).
Thus, measuring of the mean value of spin allows to find experimentally the value of entanglement of spin with other quantum system in pure state.
 It is worth mentioning that spin is maximally entangled with other quantum system when its mean value is zero, $|\left<\bm{\sigma}\right>|=0$, and it is separable when $|\left<\bm{\sigma}\right>|=1$ that follows from (\ref{EntSpin}).

We have also considered entanglement of two spins in mixed states of rank-2
which are defined on subspaces spanned by vectors
$|\uparrow\downarrow\rangle$, $|\downarrow\uparrow\rangle$ or
subspace spanned by vectors $|\uparrow\uparrow\rangle$,
$|\downarrow\downarrow\rangle$. For these cases we have found
explicit expression for geometric entanglement and have shown,
that the geometric entanglement can be expressed by the mean
values of spin correlations (\ref{Ecorrel}). This result allows to
find experimentally geometric entanglement by measuring
corresponding mean value of correlations of two spins in mixed
state. Using our result (\ref{Erhoa}) we have concluded that the
maximally entangled states with $E=1/2$ can be obtained only in
the case of $|{\bf a}|=1$, which corresponds to pure states. For
arbitrary mixed state $|{\bf a}|<1$, thus, the magnitude of
entanglement is less than the maximal value $1/2$. From
(\ref{Ecorrel}) we have concluded that in the case of
$<\sigma_1^x\sigma_2^x>=0$ and $<\sigma_1^y\sigma_2^x>=0$ the
two-spin mixed states are non-entangled.
These results are also generalized on rank-2 mixed system of arbitrary number of spins.

Our results (\ref{EntSpin}), (\ref{Ecorrel}) connect the
entanglement with observables. Therefore, the present approach
provides the effective way of an experimental determination of
geometric measure of entanglement for considered pure and mixed
states of rank-2.

The relation of the entanglement with the mean value of spin
(\ref{EntSpin}) is very useful for the calculation of
entanglement. As an example we consider the entanglement of the
first spin with others in spin chain during the evolution with the
Ising Hamiltonian. For the calculation of entanglement it is not
necessary to find state vector during the evolution explicitly. It
is enough to find mean value of the first spin and with the help
of (\ref{EntSpin}) to find entanglement. In such a way we find in
explicit form the geometric entanglement of the first spin with
others (\ref{Echain}) in the Ising spin chain during the
evolution.

We also show the usefulness of formula (\ref{Erhoa}) for
calculation geometric entanglement of two spins in mixed state.
For this purpose we consider ensemble of two-spin systems
described by Ising Hamiltonian and placed in transferse
fluctuating magnetic field. Using (\ref{Erhoa}) we find geometric
entanglement for this system in explicit form (\ref{Ent1}).
As another example we consider the Schr\"odinger cat quantum state of $N$ particles
and find geometric entanglement of one spin with others during the evolution
and decoherence of this system.

Finally let us show that our explicit expression for geometric entanglement (\ref{Erhoa})
reproduce results presented in \cite{Boy16}. Let us consider one of them presented on
Fig.2 in \cite{Boy16}. Namely, the authors state that the Bloch sphere has only one
line of separable states: all the states along the line connecting $|00\rangle$ and $|11\rangle$
(in our notation $|\uparrow\uparrow\rangle$ and $|\downarrow\downarrow\rangle$)
are separable. This result immediately follows from our result (\ref{Erhoa}). Really, the geometric entanglement
$E(\rho)=0$ when $a_x=a_y=0$ that corresponds to the line connecting
$|\uparrow\uparrow\rangle$ and $|\downarrow\downarrow\rangle$ on the Bloch sphere.
Advantage of our result is that we have explicit expression (\ref{Erhoa}) for geometric entanglement
of rank-2 mixed states and which is suitable for arbitrary number of spins. This give a possibility to
calculate the value of geometric entanglement for different quantum systems that was demonstrated in our
paper.

\section*{Appendix}
The question of finding maximum (\ref{propl}) can be reformulated in geometrical terms. The constraint (\ref{const}) can be interpreted in such a way, that we have an inextensible cord of unit length.
The ends of this cord, according to (\ref{cond}), are placed at the beginning and at the end of the vector ${\bf a}$.
The problem of finding  maximum (\ref{propl}) corresponds to the pulling of the cord in such a way that the projection of
$K'$  on the $Z$-axis  is maximal (see fig. 1).
\begin{figure}[h]
\begin{center}
\includegraphics[width=7cm]{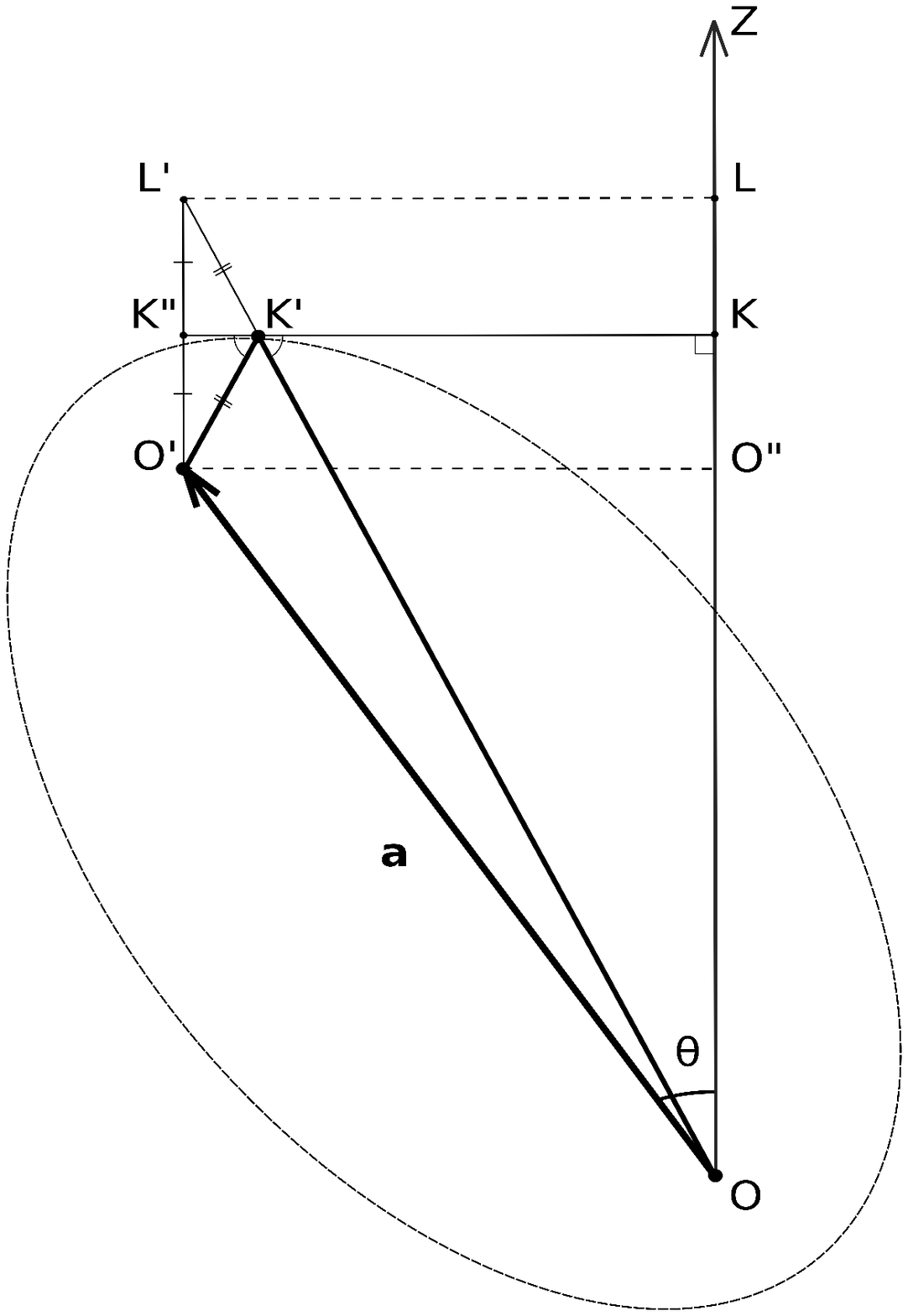}
\end{center}
\caption{Geometric solution of problem (\ref{propl}).}
\label{fig.1}
\end{figure}
The quantity $\max\sum_i |l_i^z|$ is equal to $|OK|+|O'K''|$.
The point $K'$ lays on ellipse, because $|OK'|+|K'O'|=1$ as fixed length of cord. The focuses of the ellipse are pointed at $O$ and $O'$. The tangent to the ellipse at point $K'$ is perpendicular to the $Z$-axis. Let us make the construction as is shown in fig.1.
We continue line $OK'$ to the point $L'$, where $L'$ lays on $O'L'$, which is parallel to the $Z$-axis.
Let as show that right triangles $\triangle O'K''K'$ and $\triangle L'K'K''$ are equal.
It is known that, if a rays' source is placed at one focus of an elliptic mirror, all rays on the plane of the ellipse are reflected to the second focus. This means that the angles $\widehat{O'K'K''}$ and $\widehat{OK'K}$ are equal. According to the construction the angles $\widehat{OK'K}$ and $\widehat{L'K'K''}$ are equal as vertical. Therefore, the angle $\widehat{L'K'K''}$  equals to $\widehat{O'K'K''}$. Thus, right triangles $\triangle O'K''K'$ and $\triangle L'K'K''$ have equal angles adjacent to the common side. As the consequence, the said triangles are equal. So, the sought quantity reads $\max\sum_i |l_i^z|=|OK|+|K'L'|=|OL|$.  We can find $|OL|$ from the right triangle $\triangle (OL'L)$ as $|OL|=\sqrt{|OL'|^2-|LL'|^2}$. The hypotenuse of triangle $\triangle (OL'L)$ equals $|OL'|= |OK'|+|K'L'|=|OK'|+|K'O'|=1$ as  the length of the cord.
 The leg of this triangle is the following $|LL'|=|O'O''|=a\sin\theta$, where $a=|{\bf a}|$, $\theta$ is the angle between $\bf a$ and the $Z$-axis.
Finally, we have
\begin{eqnarray}
|OL|=\sqrt{1-a^2\sin^2\theta}.
\end{eqnarray}


\begin{thebibliography}{99}
\bibitem{Hor09} R. Horodecki, P. Horodecki, M. Horodecki, K. Horodecki, Rev. Mod. Phys. {\bf 81}, 865 (2009)
\bibitem{Ple07} M. B. Plenio, S. Virmani, Quantum Inf. Comp. {\bf 7}, 1 (2007)
\bibitem{Shi95} A. Shimony, Ann. N. Y. Acad. Sci. {\bf 755}, 675 (1995)
\bibitem{BroHu01} D. C. Brody, L.P. Hughston, J. Geom. Phys. 38, 19-53 (2001)
\bibitem{Wei03} T. C. Wei, P. M. Goldbart, Phys. Rev. A {\bf 68}, 042307 (2003)
\bibitem{Chen14} L. Chen, M. Aulbach, M. Hajdu\v{s}ek, Phys. Rev. A {\bf 89}, 042305 (2014)
\bibitem{Guh02} O. Guhne, P. Hyllus, D. Bruss A. Ekert, et al., Phys. Rev. A {\bf 66}, 062305 (2002)
\bibitem{Alt05} J. B. Altepeter, E. R. Jeffrey, P. G. Kwiat, S. Tanzilli, N. Gisin, and A. Ac\'\i{}n, Phys. Rev. Lett. {\bf 95}, 033601 (2005)
\bibitem{Kot07} Ch. Kothe, G. Bj\"ork, Phys. Rev. A {\bf 75}, 012336 (2007)
\bibitem{Wal07} S. P. Walborn, P. H. Souto Ribeiro, L. Davidovich, F. Mintert, and A. Buchleitner,
Phys. Rev. A {\bf 75}, 032338 (2007)
\bibitem{Enk07} S. J. van Enk, N. L\"utkenhaus, H. J. Kimble, Phys. Rev. A {\bf 75}, 052318 (2007)
\bibitem{Fei09} Shao-Ming Fei, Ming-Jing Zhao, Kai Chen, Zhi-Xi Wang, Phys. Rev. A {\bf 80}, 032320 (2009)
\bibitem{Bri10} G. Brida, I. P. Degiovanni, A. Florio, et al, Phys. Rev. Lett. {\bf 104}, 100501 (2010)
\bibitem{Law14} T. Lawson, A. Pappa, B. Bourdoncle, et al. Phys. Rev. A {\bf 90}, 042336 (2014)
\bibitem{Dai14} Jibo Dai, Yink Loong Len, Yong Siah Teo, Berthold-Georg Englert, Leonid A. Krivitsky,
Phys. Rev. Lett. {\bf 113}, 170402 (2014)
\bibitem{Bar15} K. Bartkiewicz, J. Beran, K. Lemr, M. Norek, and A. Miranowicz,
Phys. Rev. A {\bf 91}, 022323 (2015)
\bibitem{Boy16}  Michel Boyer, Rotem Liss, Tal Mor, On the Geometry of Entanglement, arXiv:1608.00994
\bibitem{Sen10} A. Sen(De), U. Sen, Phys. Rev. A {\bf 81}, 012308 (2010)
\end{thebibliography}
\end{document}